# The New Physics of Cosmic Redshift

Wolfgang HEBEL *

## Abstract
Light propagates straight from the source to the receiver according to the rules of classical optics. Light rays received on Earth from distant stars show redshift, being attributed conventionally to the well-known Doppler-effect of wave dynamics. The present study concludes that cosmic redshift rather is an effect of the quantum mechanical propagation of photons as explained by Nobel Laureate Richard FEYNMAN in his book on QED [2]. This alternative physics of cosmic redshift is fundamentally different from the conventional velocity argument and can therefore do without the controversial big bang idea.

## Conventional theory of cosmic redshift

In 1929, Edwin HUBBLE discovered that light from distant stars exhibits longer wavelengths or reduced frequencies than that from similar radiation sources on Earth. All specific frequency lines in the electromagnetic spectrum of distant cosmic radiation sources appeared shifted towards the red end of the visible spectrum. His discovery therefore was called *astronomic redshift* defined by the redshift ratio, $z = \Delta l / l_o$ (1), of which $\Delta l$ means the elongation of a specific wave length and $l_o$ the original wave length of emission at the source. Frequency n and wave length l of the radiation are correlated by the velocity of light in vacuum, $c = n \times l$ (2).

From on its discovery, the astronomic or cosmic redshift was attributed principally to the well-known Doppler-effect of wave dynamics, manifest when source and receiver of waves move relative to each other. HUBBLE's discovery, therefore, proved as it were that all distant stars (galaxies) were escaping from Earth, i.e. that our Universe was expanding. The apparent radial recession velocity away from Earth can be calculated following the Doppler equation :
$n/n_o = (1-v/c)$ (3), of which n means a specific frequency of the arriving light, $n_o$ the corresponding frequency of emission at the source, v the recession velocity of the source and c the velocity of light in vacuum. Simultaneously, HUBBLE discovered that the apparent recession velocity of cosmic radiation sources increases proportionally to their distance r from Earth: $v = H \times r$ (4), H means the famous HUBBLE-constant currently estimated at about 70 km/s per megaparsec or per 3.26 million light-years. HUBBLE's discoveries soon led to our current view of the Universe, assuming that it came into existence by a gigantic explosion, the so-called Big Bang, which spontaneously arose from a tiny volume of unimaginable high temperature followed up by adiabatic expansion and the condensation of matter while cooling down. Apparently, this expansion process is still going on today. Following HUBBLE's law (4), the escape velocity of an extremely distant galaxy might gain ultimately the velocity of light in vacuum, meaning its redshift ratio would theoretically equal unity. Physically of course, this is impossible nevertheless redshift ratios of z=5 and even z=7 have been measured in recent years showing supernova explosions, which apparently occurred further away from Earth than the age of the Universe postulated at about 14 billion light-years. In addition, as well known, various other inconsistencies weigh upon the Big Bang theory and many a scientist therefore questions this view.

*Ex-EU Scientific Coordinator
Brussels, Belgium, 30.12.2009
Email: wolfgang.hebel@telenet.be

# New physics of cosmic redshift

Richard FEYNMAN received the Nobel Prize of physics in 1965, for his pioneering studies on *quantum electro-dynamics*, explaining among other things the interactions of photons with matter. In his book on "QED – *The Strange Theory of Light and Matter*" published in 1985 [2], he explained also the quantum mechanics of the linear propagation of photons. He showed that from the countless photons being emitted into all directions from a radiation source, only those are really effective, which travel the distance to a given receiver in straight line and in close company i.e. when they travel this distance from the source to the receiver within the shortest possible time. All other photons taking different, dispersed paths need more time and are therefore ineffective.

A straight stream of photons traveling from a given cosmic radiation source to the Earth will meet countless celestial bodies before arriving on Earth. These bodies as stars, planets, comets, meteorites, grains, etc. are swirling around in the Universe at typical velocities of some hundred kilometers per second and are impermeable to photons. When those bodies cross a straight ray of light, this ray will be interrupted shortly and all photons dropping onto the bodies will be removed from the stream independently of their individual energy or frequency. The photons coming thereafter will travel further and after countless interruptions only the remaining photons of the stream will arrive with the observer on Earth. He will remark that all typical spectral lines of this ray of light exhibit lower frequencies than ordinary, because the knocked-out photons did not show up in time. The ray lost a good deal of its original photons and shows a redshift ratio, which is proportional to its traveling distance through the Universe. In other words, the new redshift physics confirms HUBBLE's discovery establishing that the cosmic redshift ratio indicates how far a distant star is away from Earth.

HUBBLE's law still implies another connection. When replacing in equation (4) the recession velocity v by the product $z \times c$, i.e. by a fraction of the ultimate velocity of light, an interesting correlation arises: $z = H/c \times r$ (5). The constant factor $H/c$ can be regarded as a modified HUBBLE-constant, which amounts to 0.00023 per megaparsec or 0.00007 per million light-years. This modified HUBBLE-constant signifies the loss of photons suffered by a beam of light, which has traveled one million light-years through the Universe. It is a very small loss of photons in fact over such big distance, confirming our experience that the Universe is largely empty of solid matter. The reciprocal of the modified HUBBLE-constant i.e. 1/0.00007 gives 14 billion light-years, in accordance with the postulated age of our Universe following the conventional theory. However, the meaning is different. It shows namely that ordinary starlight cannot propagate further through the Universe than 14 billion light-years. All photons grouped in a straight light beam from a remote source to the Earth would have got lost due to the absorption effect of celestial matter. Usually, we cannot look deeper into the Universe than this distance, corresponding theoretically to a redshift ratio of unity. However, what about those bigger redshift ratios of $z=5$ or even $z=7$, which have been observed in recent time?

We know from nuclear physics that energetic radiation penetrating through a shielding medium will be absorbed according to the general correlation, $E = E_o \times e^{-\mu r}$ (6), E means the radiation energy behind the shield, $E_o$ the energy of emission at the source, $\mu$ the absorption coefficient of the shielding medium and r the traveling distance through the medium. In the present case, it makes sense to take the modified HUBBLE-constant $H/c$ as a cosmic photon absorption coefficient and r as traveling distance of a stream of photons through the Universe.

The energy of photons emitted by a heat source corresponds to the well-known correlation,
$E = k \times T$ (7), T means the surface temperature of the source and k is the BOLTZMANN-constant. On the other hand, the photon energy corresponds to PLANCK's equation,

*Ex-EU Scientific Coordinator
Brussels, Belgium, 30.12.2009
Email: wolfgang.hebel@telenet.be

3$E = h \times n$ (8), h is the PLANCK-constant and n the frequency of the photons. From equations (6), (7), and (8) follow the correlations, $n/n_o = T/T_o$ and $E/E_o = e^{-H/c \cdot r}$ (9).

Referring to the previous equations (1) and (2), one finds that the cosmic redshift ratio equals $z = e^{H/c \cdot r} - 1$ (10).

The distance of a cosmic radiation source therefore can be estimated from its redshift ratio according to $r = 1/H/c \times \ln(z+1)$ (11), with $H/c = 0.00007$ per million light-year. For a redshift ratio of z=1, for instance, one finds 9.9 billion light-years, and for z=5 a cosmic distance of 25.6 billion light-years.

## Conclusion

In contrast to the traditional redshift theory, the present new physics of cosmic redshift doesn't show any upper limit of 'z' or any restricted age of our Universe respectively. Referring to the aforementioned equations, the cosmic redshift ratio also equals, $z = T_o/T - 1$ (12), i.e. it is proportional to the ratio of the surface temperature of a cosmic radiation source $T_o$ and its apparent temperature T observed on Earth. The light rays from a remote cluster of galaxies at an average surface temperature of about 5800°K as our sun would show the apparent temperature of 970°K (700°C) on Earth, when arriving from a cosmic distance of z=5 or 25.6 billion light-years. This cluster, of which most of the photons got lost on the way to Earth, would be invisible to ordinary optical telescopes. However, cosmic radiation sources of much higher surface temperatures or bigger emission energy like supernova explosions would still be visible over such extraordinary distances, which largely exceed the age of our Universe as postulated from the big-bang hypothesis.

In contrast to the conventional theory of cosmic redshift, the present new physics does not present any difficulty to explain those observations. In addition, this physics still offers another interesting conclusion. The phenomenon of cosmic microwave background radiation (CMB) namely, can be explained as the thermodynamic background radiation of a Universe without frontiers. For, this ubiquitous thermodynamic background radiation noise suggests that innumerable stars still exist in the remote Universe far beyond the practical limits of detecting individual radiation sources.

The present revelations on a new physics of cosmic redshift were made possible principally thanks to the findings of Richard FEYNMAN described in his book on "QED - *The Strange Theory of Light and Matter*". His quantum mechanics findings were not yet known to Edwin HUBBLE previously when he made his seminal discoveries on cosmic redshift. Nevertheless, Hubble always remained cautious against the importunate idea of a big-bang and an expanding Universe to explain his research results.

R e f e r e n c e s :
[1] Paul Davies: *The New Physics*
    Cambridge University Press, New York, 1989
[2] Richard P. Feynman: "QED – The Strange Theory of Light and Matter"
    Princeton University Press, Princeton, 1985
[3] Craig J. Hogan: *Revolution in Cosmology*
    Scientific American, p. 27-49, January 1999
[4] Ann Finkbeiner (ed): *Seeing the Universe's red dawn*
    SCIENCE, p. 392, 16 Oct. 1998
[5] Floyd E. Bloom: *Breakthroughs 1998*
    SCIENCE, p. 2193, 18 Dec. 1998
[6] Wolfgang Hebel: *The Mystery of Life – Does Science hold the Key?*
    German University Press (GUP), Baden-Baden, 2007
*Ex-EU Scientific Coordinator
Brussels, Belgium, 30.12.2009
Email: wolfgang.hebel@telenet.be

*Ex-EU Scientific Coordinator
Brussels, Belgium, 30.12.2009
Email: wolfgang.hebel@telenet.be